\documentclass[preprint]{elsarticle}

\usepackage[utf8]{inputenc}  
\usepackage[T1]{fontenc}     
\usepackage{lmodern}         
\usepackage{microtype}       
\usepackage[scaled]{helvet} 
\usepackage[english]{babel}  
\usepackage{graphicx}        
\usepackage[bookmarks=false]{hyperref}            
\usepackage{comment}         
\usepackage{booktabs}
\usepackage{makecell}
\usepackage{tabularx}
\usepackage{multirow}
\newcolumntype{L}{>{\raggedright\arraybackslash}X}
\biboptions{sort&compress}   

\usepackage{amsmath,amsfonts,amssymb}

\usepackage{subfig}

\usepackage{xcolor}
\usepackage{todonotes}

\graphicspath{{fig/}}

\begin{document}

\begin{frontmatter}

\title{EMPD: An Event-based Multimodal Physiological Dataset for Remote Pulse Wave Detection}

\author{Qian Feng}
\author{Pengfei Li}
\author{Rongshan Gao}
\author{Jiale Xu}
\author{Rui Gong}
\author{Yidi Li\corref{cor1}}
\address{College of Computer Science and Technology, Taiyuan University of Technology
}
\cortext[cor1]{Corresponding author: liyidi@tyut.edu.cn}

\begin{abstract}
Remote photoplethysmography (rPPG) based on traditional frame-based cameras often struggles with motion artifacts and limited temporal resolution. To address these limitations, we introduce EMPD (Event-based Multimodal Physiological Dataset), the first benchmark dataset specifically designed for non-contact physiological sensing via event cameras. The dataset leverages a laser-assisted acquisition system where a high-coherence laser modulates subtle skin vibrations from the radial artery into significant signals detectable by a neuromorphic sensor.
The hardware platform integrates a high-resolution event camera to capture micro-motions and intensity transients, an industrial RGB camera to provide traditional rPPG benchmarks, and a clinical-grade pulse oximeter to record ground truth PPG waveforms. EMPD contains 193 valid records collected from 83 subjects, covering a wide heart rate range (40–110 BPM) under both resting and post-exercise conditions. By providing precisely synchronized multimodal data with microsecond-level temporal precision, EMPD serves as a crucial resource for developing robust algorithms in the field of neuromorphic physiological monitoring.
The dataset is publicly available at: \url{https://doi.org/10.5281/zenodo.18765701}

\end{abstract}
\begin{keyword}
Event-based dataset \sep Multimodal sensing \sep Physiological signals \sep Remote photoplethysmography
\end{keyword}
\end{frontmatter}


\section{Introduction} 
Remote photoplethysmography (rPPG) has emerged as a compelling non-contact technique for continuous physiological monitoring \cite{xiao2024remote, saikevivcius2025advancements}. Public benchmark datasets, such as PURE \cite{stricker2014non}, UBFC-RPPG \cite{bobbia2019unsupervised}, and MMPD \cite{tang2023mmpd}, have driven significant advancements in rPPG algorithms by providing diverse scenarios ranging from controlled head motions to mobile applications. However, existing mainstream datasets are fundamentally constrained by their reliance on traditional frame-based camera acquisition. These conventional sensors typically operate at fixed, relatively low frame rates (e.g., 30–60 fps) and suffer from exposure-induced motion blur, making them highly susceptible to motion artifacts and under-sampling when capturing the high-frequency micro-motions of cardiovascular activities \cite{gallegoEventBasedVisionSurvey2022, jagtapHeartRateDetection2023, lvStructuralVibrationFrequency2024}.

To address these inherent hardware bottlenecks, event cameras present a paradigm-shifting alternative. Unlike traditional cameras that capture full images at constant intervals \cite{gallegoEventBasedVisionSurvey2022}, event cameras respond asynchronously to pixel-level brightness changes, offering microsecond-level temporal resolution and a remarkably high dynamic range. These unique characteristics inherently compensate for the limitations of frame-based sensors, effectively mitigating motion blur and capturing ultra-fast physiological transients \cite{jagtapHeartRateDetection2023}.

Furthermore, because event cameras output sparse, asynchronous event streams rather than dense, synchronous frames \cite{debraine2019stereo}, their integration necessitates the application of novel neuromorphic vision techniques. These techniques are specifically designed to process event-based data efficiently, opening new avenues for robust, low-latency pulse wave analysis. However, the development and evaluation of such neuromorphic physiological sensing algorithms are currently hindered by a critical lack of dedicated, high-quality benchmark datasets.

To bridge this gap, we introduce the Event-based Multimodal Physiological Dataset (EMPD), the first benchmark specifically constructed for non-contact physiological sensing utilizing event cameras. To overcome the challenge of detecting low-contrast physiological micro-motions, we engineered a laser-assisted data acquisition platform. This system utilizes a high-coherence 660nm laser as a detection medium, modulating the subtle, pulse-induced skin vibrations of the radial artery into high-contrast intensity changes that are easily perceivable by the event camera.

By providing a rigorously synchronized, high-spatiotemporal resolution modality, EMPD serves as a crucial complement to existing frame-based datasets, aiming to catalyze the innovation of neuromorphic biomedical signal processing.

In summary, the main contributions of this work are as follows:
\begin{itemize}
    \item EMPD is the pioneering dataset that introduces event cameras to the remote physiological measurement task.
    \item The dataset provides precisely aligned, microsecond-level event streams alongside industrial-grade RGB videos, achieved through a customized synchronization framework.
    \item We include synchronously recorded contact PPG signals from a clinical-grade pulse oximeter as the definitive gold standard for evaluation.
    \item Comprising 193 valid recordings from 83 subjects, the dataset encompasses a broad heart rate spectrum across both resting and post-exercise states, ensuring robust algorithmic validation.
    
\end{itemize}

\section{Related Work} 
Public benchmark datasets have played a critical role in advancing the development and performance evaluation of remote photoplethysmography (rPPG) algorithms. As summarized in Table \ref{tab:dataset_comparison}, we review five representative datasets in this field, comparing them across subject scale, sensor modalities, total duration, and scene diversity.

\begin{table}[!t]
    \centering
    \caption{Comparison between representative PPG datasets and our proposed EMPD dataset.}
    \label{tab:dataset_comparison}
    
    \footnotesize 
    
    \setlength{\tabcolsep}{3pt} 
    
    \begin{tabularx}{\textwidth}{@{} l c c L c L @{}}
        \toprule
        \textbf{Datasets} & \textbf{Year} & \textbf{Subjects} & \makecell{\textbf{Sensor} \\ \textbf{modality}} & \makecell{\textbf{Total} \\ \textbf{duration}} & \makecell{\textbf{Scene} \\ \textbf{diversity}} \\
        \midrule
        \textbf{PURE} \cite{stricker2014non} & 2014 & 10 & RGB Camera, Pulse Oximeter & 60min & Head motions, Talk, Lighting \\
        \textbf{UBFC-RPPG} \cite{bobbia2019unsupervised} & 2019 & 42 & RGB  Webcam, Pulse Oximeter & 70min & Math game, Stress induction \\
        \textbf{OBF} \cite{li2018obf} & 2018 & 106 & RGB \& NIR Camera, ECG & 1000min & Atrial fibrillation, Exercise \\
        \textbf{MMPD} \cite{tang2023mmpd} & 2023 & 33 & Mobile Camera, Pulse Oximeter & 660min & Mobile, Multi-lighting, Exercise \\
        \textbf{MSPM} \cite{spethMSPMMultisitePhysiological2024} & 2024 & 103 & RGB, NIR, Thermal, Contact PPG & 1480min & Multi-site, BP, Breath hold \\
        \midrule
        \textbf{EMPD (Ours)} & 2025 & 83 &  Event \& RGB Camera, Pulse Oximeter & 219min & Rest, Exercise\\
        \bottomrule
    \end{tabularx}
\end{table}
\subsection{Traditional Frame-based rPPG Datasets}
Early dataset collection efforts primarily focused on controlled laboratory environments to isolate specific physiological or motion-related variables. The PURE \cite{stricker2014non} dataset was specifically designed to provide a controlled baseline for evaluating the impact of head motions on remote physiological measurements. It comprises 10 subjects who performed six specific motion sequences under natural lighting, ranging from complete rest and talking to various degrees of head translation and rotation. The video frames were captured using an industrial RGB camera at 30 fps and stored in a lossless PNG format to guarantee high image quality. However, the dataset's relatively small scale, simple environmental settings, and limited amplitude of heart rate variations restrict its utility for evaluating algorithms in highly dynamic scenarios.

To introduce more natural physiological fluctuations, the UBFC-RPPG \cite{bobbia2019unsupervised} dataset required its 42 subjects to participate in a time-constrained mathematical game. This experimental design successfully induced natural heart rate variations without relying solely on physical movement, thereby simulating realistic human-computer interaction scenarios and increasing data diversity. The videos were recorded indoors using a Logitech C920 webcam in an uncompressed 8-bit RGB format. Despite these advantages, the dataset features a largely homogenous distribution of skin tones and a relatively small number of video clips, which to some extent limits its capacity to train highly complex, data-driven algorithms.

Subsequent benchmarks aimed to scale up the data volume and introduce clinical relevance. The OBF \cite{li2018obf} dataset was introduced to support deeper physiological signal analysis, recruiting 100 healthy subjects for resting and post-exercise testing, alongside a unique cohort of patients with atrial fibrillation (AF). Utilizing synchronized dual-modality RGB and near-infrared (NIR) cameras at a high frame rate of 60 fps, OBF \cite{li2018obf} provides approximately 1,000 minutes of high-quality video recordings. Nevertheless, the dataset was collected under highly stable lighting conditions, and the subjects exhibited minimal body movement, which does not fully represent the complexities of real-world monitoring.

Focusing on the pervasive mobile health sector, the MMPD \cite{tang2023mmpd} dataset specifically addressed the scarcity of data collected via mobile devices and the lack of skin tone diversity. It contains 11 hours of video data from 33 subjects, captured entirely using a Samsung Galaxy S22 Ultra smartphone. MMPD \cite{tang2023mmpd} is highly challenging, covering four complex lighting environments and four distinct motion patterns. However, to facilitate data sharing and transmission, the videos underwent compression processing, which inevitably degrades the subtle pixel-level intensity variations crucial for accurate rPPG extraction. Recently, the MSPM \cite{spethMSPMMultisitePhysiological2024} dataset expanded the research scope to large-scale, full-body physiological monitoring. Diverging from traditional face-centric datasets, MSPM \cite{spethMSPMMultisitePhysiological2024} captured multi-view RGB, NIR, and thermal videos from 103 subjects, while synchronously recording contact PPG signals from 10 different body sites and cuff blood pressure data.

\subsection{The Need for Neuromorphic Benchmarks}
In summary, all existing mainstream datasets fundamentally rely on traditional frame-based camera acquisition. Consequently, they universally lack the microsecond-level temporal resolution and high dynamic range that are intrinsic characteristics of event cameras. This architectural hardware limitation heavily restricts researchers from leveraging emerging neuromorphic vision technologies to overcome persistent bottlenecks in rPPG, such as exposure-induced motion blur and temporal under-sampling during high-frequency physiological events. To bridge this critical gap, we introduce the EMPD dataset, aiming to provide a novel, high-spatiotemporal-resolution modality reference for non-contact physiological monitoring that perfectly complements existing frame-based benchmarks.

\section{The EMPD Dataset}
To evaluate the pulse wave detection method and propel the development of neuromorphic physiological sensing, we constructed the Event-based Multimodal Physiological Dataset (EMPD). To the best of our knowledge, this is the first non-contact multimodal physiological signal benchmark dataset specifically developed for event cameras in the academic community.

\begin{figure}[!t]
    \centering
    \includegraphics[width=1.0\textwidth]{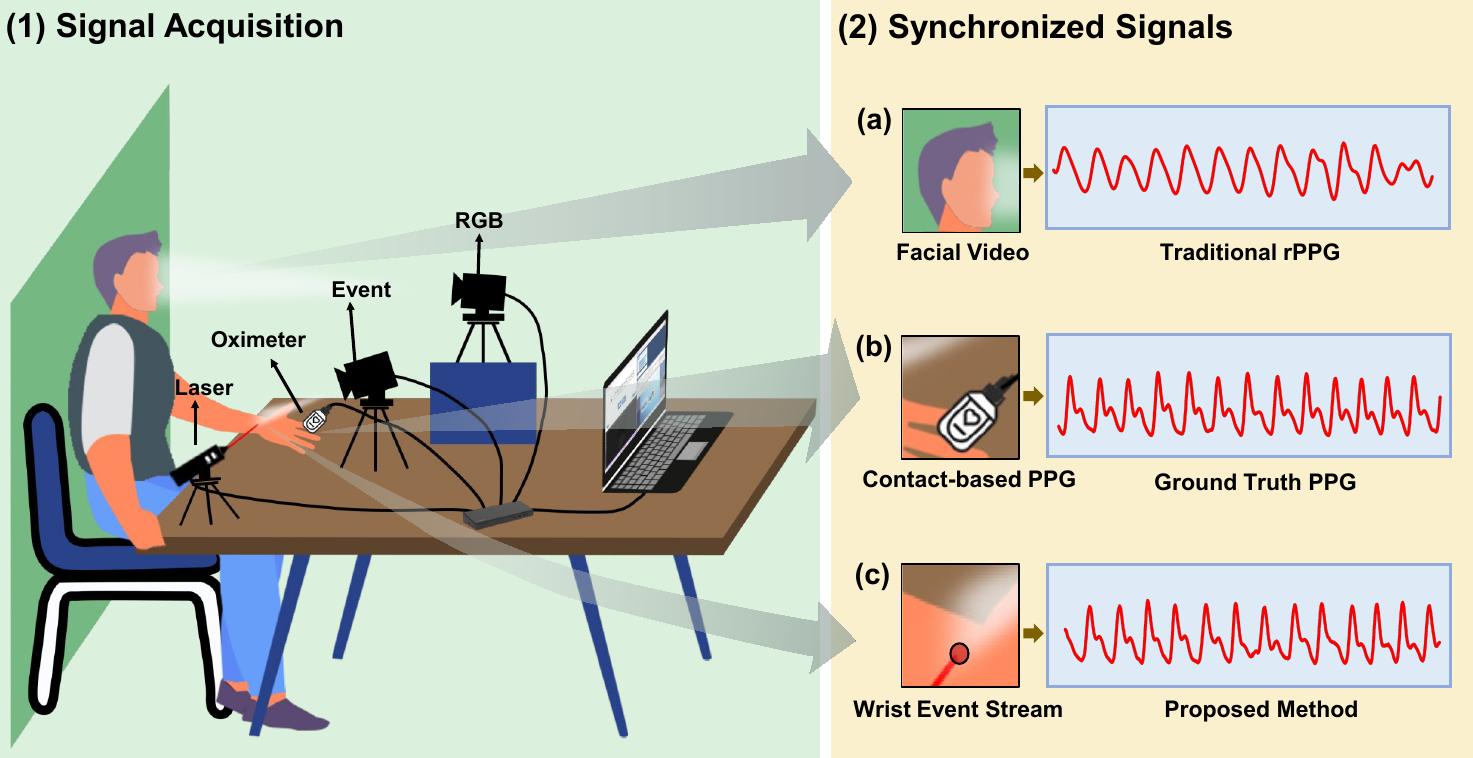}
    \caption{\textbf{Schematic of the pulse signal acquisition system.} The system simultaneously records three signal sources: (a) facial video for traditional remote photoplethysmography (rPPG) extraction, (b) contact-based PPG as the ground truth, and (c) wrist event stream for our proposed method.}
    \label{fig:system_schematic}
\end{figure}

\begin{figure}[!b]
    \centering
    \includegraphics[width=0.4\textwidth]{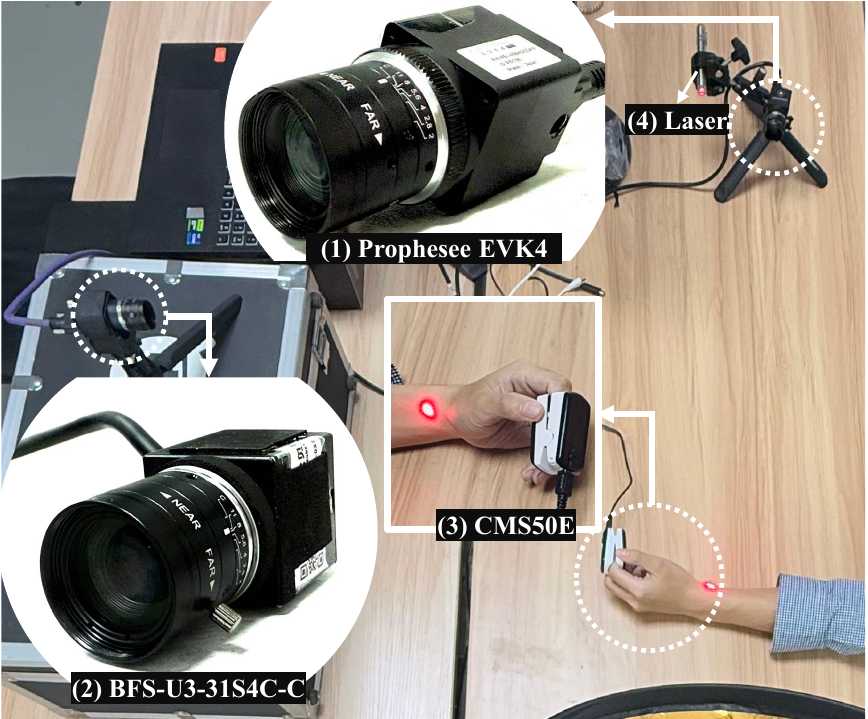}
    \caption{Real-world picture of the experimental equipment. It includes: (1) Prophesee EVK4 event camera; (2) FLIR industrial-grade RGB camera; (3) CONTEC fingertip pulse oximeter; (4) Active laser light source used to enhance micro-motion signals.}
    \label{fig:equipment}
\end{figure}

\subsection{Laser-Assisted Data Acquisition Platform}
We designed and implemented an integrated laser-assisted multimodal pulse signal acquisition platform. To effectively capture the subtle skin vibrations of the radial artery, a high-coherence laser is employed as a detection medium, modulating the micro-motions into significant signals perceivable by the event camera. The hardware platform integrates a neuromorphic vision sensor with traditional optical and clinical-grade sensors. A customized synchronization control framework eliminates temporal drift among the heterogeneous sensors, ensuring precise alignment between the asynchronous event streams and the physiological ground-truth waveforms. The overall system architecture is illustrated in Fig. \ref{fig:system_schematic}.

The core hardware components of the acquisition system (shown in Fig. \ref{fig:equipment}) include:
\begin{itemize}
    \item Event Camera (Prophesee EVK4): Serving as the core sensor, it captures micro-motions and transient light intensity changes caused by pulse pulsations at the wrist's radial artery. It operates at a spatial resolution of $1280 \times 720$ with a timestamp precision of 1 $\mu s$, enabling the capture of high-frequency physiological features imperceptible to conventional cameras.
    \item Industrial RGB Camera (FLIR BFS-U3-31S4C-C): Positioned approximately 1.5 m from the subject's face, it synchronously records facial videos at a resolution of $2048 \times 1536$ (downsampled to $640 \times 480$ in the release version) at 30 Hz. This camera provides traditional rPPG benchmark data for fair comparative evaluations.
    \item Contact Pulse Oximeter (CONTEC CMS50E): Worn on the subject's fingertip, this device continuously acquires ground-truth PPG waveforms and heart rate data at a sampling rate of 60 Hz.
    \item Active Light Source: A 660 nm laser pointer projects a light spot onto the radial artery area. This active illumination enhances the contrast variations on the skin surface, significantly increasing the event camera's sensitivity to pulse-induced micro-vibrations.
\end{itemize}

\subsection{Experimental Protocol}
Data collection was conducted in an indoor environment characterized by stable lighting and minimal acoustic interference. Subjects maintained a comfortable sitting posture while keeping their left wrist completely stationary. During the acquisition process, the event camera was positioned approximately 40 cm from the wrist, and the RGB camera was placed 1.5 m from the face, both meticulously focused. After the subject wore the pulse oximeter and the point of the strongest pulsation on the radial artery was calibrated, the system initiated synchronous recording.

\subsection{Dataset Characteristics}
The EMPD dataset recruited a total of 83 healthy adult volunteers (57 males and 26 females) with ages ranging from 19 to 25 years. To ensure comprehensive coverage of diverse physiological states, a subset of subjects performed deep squats (10 repetitions) prior to recording to intentionally elevate their heart rates, yielding a total of 193 continuous physiological recordings. 
To capture the dynamic temporal characteristics of the physiological signals and standardize the input length for algorithmic evaluation, a sliding window approach (5-second window, 1-second stride) was applied to the continuous recordings. Consequently, the finalized dataset yields a total of 7,527 valid data segments, partitioned into 1,404 resting-state samples and 6,123 post-exercise samples.

\begin{figure}[t]
    \centering
    \includegraphics[width=1.0\textwidth]{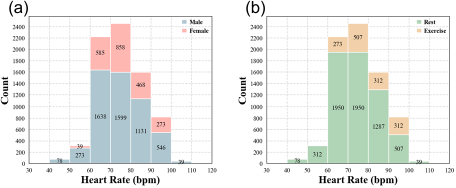}
    \caption{Histogram of heart rate (BPM) distribution measured by the contact pulse oximeter across 7527 valid samples. Data are stratified by (a) gender (Male vs. Female) and (b) activity state (Rest vs. Exercise).}  
    \label{fig:distribution}
\end{figure}

Statistical analysis of the heart rate data extracted from all valid records reveals a quasi-normal distribution, as depicted in Fig. \ref{fig:distribution}. The heart rates are predominantly concentrated within the 60–90 BPM range, effectively covering a broad physiological spectrum from 40 to 110 BPM. Figure \ref{fig:distribution}(a) illustrates the heart rate distribution stratified by gender, while Figure \ref{fig:distribution}(b) highlights the differences between resting and post-exercise states. These robust data characteristics demonstrate that the EMPD dataset possesses excellent diversity and representativeness, providing a solid foundation for evaluating algorithmic robustness under varying heart rate conditions.

\section{Experiments and Baseline Evaluation}
\subsection{Event-Driven Baseline Framework}

To demonstrate the utility of the EMPD dataset and establish a standardized benchmark for future research, we introduce an event-driven baseline framework. Since raw event cameras output asynchronous and sparse event streams rather than synchronous image frames, traditional video-based rPPG algorithms cannot be directly applied. The baseline framework addresses this modality gap through a two-stage pipeline:

\textbf{(1) Event Aggregation Module (EAM):} The first stage bridges the domain gap by converting the asynchronous event stream into a uniformly sampled, discrete-time signal sequence. It performs spatial Region of Interest (ROI) localization to isolate the radial artery area and applies temporal binning to aggregate event counts into fixed time intervals aligned with the pulse oximeter's sampling rate. A detrending operation is then applied to mitigate low-frequency baseline drift.

\textbf{(2) Waveform Reconstruction Module (WRM):} The second stage maps the aggregated one-dimensional event sequence to a high-fidelity pulse waveform. We deploy a Generative Adversarial Network (GAN) architecture. The generator utilizes a 1D U-Net to capture multi-scale temporal features, while a power spectral density (PSD) based discriminator acts in the frequency domain to ensure the synthesized waveform accurately preserves physiological rhythms and harmonic structures. 

This end-to-end mapping from raw events to physiological signals validates that the microsecond-resolution data provided in EMPD contains sufficient and extractable cardiovascular information.

\subsection{Signal Validity and Physical Plausibility}

\begin{figure}[!t]
    \centering
    \includegraphics[width=1\textwidth]{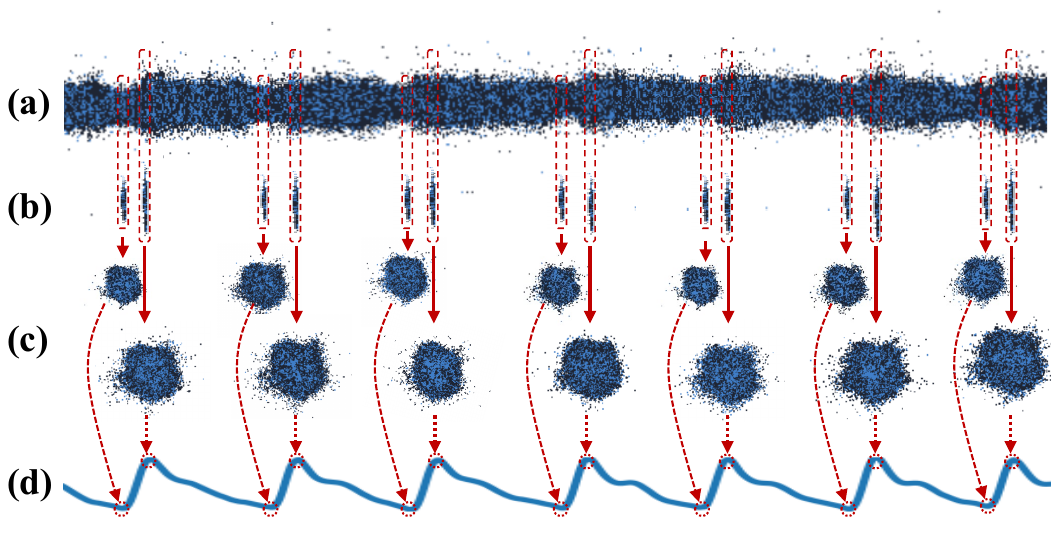} 
    \caption{\textbf{Visualization of event-pulse synchronization.} (a) Continuous raw event stream. (b) Magnified cross-sectional view of event bursts. (c) Corresponding circular cross-sections reflecting rhythmic size variations. (d) Ground-truth PPG waveform. Red dashed arrows highlight the temporal alignment between the high-density event bursts and systolic peaks. }
    \label{fig:event_sync}
\end{figure}

To validate the physical basis of our detection paradigm and demonstrate the high quality of the EMPD dataset, we visualize the temporal correspondence between the captured raw event stream and the physiological ground-truth signal. As illustrated in Fig. \ref{fig:event_sync} (top and middle panels), the raw event data manifests as distinct, rhythmic bursts of activity rather than random noise. 

Crucially, these high-density event clusters exhibit precise temporal synchronization with the systolic peaks of the reference PPG waveform (Fig. \ref{fig:event_sync}, bottom panel), as highlighted by the red dashed arrows. This strict alignment corroborates that the event camera is responding directly to the transient photometric variations induced by arterial pulsations. By confirming this correspondence, we provide a solid physical foundation proving that the raw, asynchronous event streams in EMPD contain highly valid and extractable cardiovascular information.

\subsection{Quantitative Comparison and Agreement Analysis}
To establish a comprehensive performance baseline on the EMPD dataset, we benchmarked the event-driven method against representative traditional methods (e.g., POS \cite{wang2016algorithmic}, CHROM \cite{de2013robust}, ICA \cite{pohNoncontactAutomatedCardiac2010}) and state-of-the-art (SOTA) deep learning methods, including PhysNet \cite{yu2019remote}, PhysFormer \cite{yuPhysFormerFacialVideobased2022}, and RhythmFormer \cite{zouRhythmFormerExtractingPatterned2025}. The quantitative comparison results across the entire dataset are presented in Table \ref{tab:quantitative_overall}.

As shown in the results, the event-driven baseline outperforms all frame-based methods across most evaluation metrics. Specifically, compared to the transformer-based RhythmFormer \cite{zouRhythmFormerExtractingPatterned2025} (which achieves an MAE of 3.39 BPM), the event-based approach drastically reduces the MAE to 1.18 BPM and the RMSE to 3.76 BPM. Furthermore, it achieves the highest Pearson Correlation Coefficient (PCC) of 0.94 and maintains a competitive SNR of 2.89 dB. This quantitative leap confirms that the microsecond-level temporal resolution of event streams effectively eliminates the exposure-induced motion artifacts and undersampling issues inherent in conventional RGB cameras.

\begin{table}[!t]
\centering
\caption{Performance comparison of our method with representative traditional and SOTA deep learning methods. }
\label{tab:quantitative_overall}
\footnotesize 
\setlength{\tabcolsep}{3pt} 
\begin{tabularx}{\textwidth}{lXccccc}
\toprule
\textbf{Category} & \textbf{Method} & \textbf{MAE (bpm)} & \textbf{RMSE (bpm)} & \textbf{MAPE (\%)} & \textbf{PCC} & \textbf{SNR} \\
\midrule
\multirow{7}{*}{Supervised}
& POS \cite{wang2016algorithmic} & 4.23 & 6.99 & 5.80 & 0.81 & -0.16 \\
& ICA \cite{pohNoncontactAutomatedCardiac2010} & 5.79 & 10.73 & 7.61 & 0.61 & -0.01 \\
& GREEN \cite{verkruysseRemotePlethysmographicImaging2008} & 7.29 & 13.04 & 9.30 & 0.57 & -1.44 \\
& CHROM \cite{de2013robust} & 3.57 & 5.69 & 4.88 & 0.89 & 0.44 \\
& LGI \cite{pilz2018local} & 5.87 & 10.22 & 7.61 & 0.70 & -1.93 \\
& PBV \cite{de2014improved} & 7.36 & 12.26 & 9.91 & 0.58 & -2.78 \\
& OMIT \cite{casado2023face2ppg} & 5.79 & 10.17 & 7.52 & 0.70 & -1.93 \\
\midrule
\multirow{4}{*}{Unsupervised}
& PhysNet \cite{yu2019remote} & 3.81 & 7.19 & 5.46 & 0.80 & 1.08 \\
& PhysFormer\cite{yuPhysFormerFacialVideobased2022} & 3.79 & 6.91 & 5.45 & 0.81 & 1.99 \\
& RhythmFormer\cite{zouRhythmFormerExtractingPatterned2025} & 3.39 & 5.62 & 4.47 & \ 0.91 & \textbf{3.17} \\
& \textbf{Ours} & \textbf{1.18} & \textbf{3.76} & \textbf{1.58} & \textbf{0.94} & 2.89 \\
\bottomrule
\end{tabularx}
\end{table}

\section{Conclusion and Future Work}

In this paper, we introduced EMPD, the first event-based multimodal physiological benchmark dataset specifically constructed for non-contact cardiovascular monitoring. To overcome the inherent limitations of traditional frame-based cameras—such as exposure-induced motion blur and temporal undersampling—we engineered a novel laser-assisted data acquisition paradigm. By leveraging a high-coherence laser to modulate subtle radial artery vibrations into high-contrast intensity transients, our system enables neuromorphic vision sensors to directly capture hemodynamic dynamics with microsecond-level precision.

The EMPD dataset comprises 193 synchronized, multimodal recordings from 83 healthy subjects, encompassing diverse physiological states including resting and post-exercise conditions. To demonstrate the dataset's utility, we evaluated an event-driven baseline framework alongside several state-of-the-art video-based rPPG methods. The experimental results clearly validate the physical plausibility of the event streams and confirm that the event-driven approach significantly outperforms traditional RGB-based methods in quantitative heart rate estimation accuracy.

Despite these encouraging outcomes, the current dataset was collected under relatively controlled indoor environments. Future work will focus on three main directions. First, we plan to expand the dataset to encompass more challenging, real-world scenarios, including mobile recording conditions and varying ambient illumination. Second, there is significant potential in developing more advanced, lightweight neuromorphic architectures—such as Spiking Neural Networks (SNNs)—tailored specifically for asynchronous physiological event streams. Finally, the ultra-low latency and high-fidelity nature of event-based sensing open up promising avenues for multimodal human-robot interaction. Future research could explore integrating this event-driven physiological feedback with advanced audio-visual learning mechanisms, equipping robotic systems with real-time, non-intrusive human state awareness during complex collaborative tasks. 

By making the EMPD dataset publicly available, we aim to provide a crucial foundation for the community to develop, benchmark, and accelerate innovations in neuromorphic biomedical signal processing.


\bibliographystyle{elsarticle-num}

\section*{References}

\bibliography{paper}


\end{document}